\documentclass[article]{revtex4}
\pdfoutput=1
\usepackage{color}
\usepackage{graphicx}
\usepackage{subfigure}
\usepackage{amsmath,amssymb,array}
\usepackage{enumerate}
\usepackage{hhline}
\newcommand{\mathsym}[1]{{}}

\def\10{$SO(10)$}

\newcommand{\ba}{\begin{array}}
\newcommand{\ea}{\end{array}}
\newcommand{\be}{\begin{equation}}
\newcommand{\ee}{\end{equation}}
\newcommand{\beqa}{\begin{eqnarray}}
\newcommand{\eeqa}{\end{eqnarray}}
\def\321{$SU(3)\times SU(2)\times U(1)$}
\def\b126{$\overline{126}$}

\usepackage{accents}
\usepackage{tikz}
\usepackage{lmodern}
\usetikzlibrary{decorations.pathmorphing}
\tikzset{snake it/.style={decorate, decoration=snake}}
\usepackage{fancybox}
\expandafter\ifx\csname package@font\endcsname\relax\else
\expandafter\expandafter
\expandafter\usepackage
\expandafter\expandafter
\expandafter{\csname package@font\endcsname}%
\fi
\DeclareFontFamily{OT1}{pzc}{}
\DeclareFontShape{OT1}{pzc}{m}{it}%
             {<-> s * [0.900] pzcmi7t}{}
\DeclareMathAlphabet{\mathscr}{OT1}{pzc}%
                                 {m}{it}

\usepackage{epsfig}
\usepackage{graphicx}
\usepackage{amsmath}
\usepackage{amssymb}
\usepackage{epstopdf}
\usepackage{xcolor}
\usepackage{subfigure}
\usepackage{bm}
\usepackage{dsfont}
\usepackage{hyperref}
\usepackage{braket}
\usepackage{todonotes}
\usepackage{comment}
\usepackage{tikz}
\usepackage{pgfplots}
\pgfplotsset{compat=newest}
\newcommand{\bea}{\begin{eqnarray}}
\newcommand{\eea}{\end{eqnarray}}

\begin{document}

\title{Quantum Gravity signatures in late time Universe}
\bigskip

 \author{Ankit Dhanuka}
 \email{ankitdhanuka555@gmail.com}
 \affiliation{Department of Physical Sciences, Indian Institute of Science Education \& Research (IISER) Mohali, Sector 81 SAS Nagar, Manauli PO 140306 Punjab India.}

 \author{Kinjalk Lochan}
 \email{kinjalk@iisermohali.ac.in}
 \affiliation{Department of Physical Sciences, Indian Institute of Science Education \& Research (IISER) Mohali, Sector 81 SAS Nagar, Manauli PO 140306 Punjab India.}

\begin{abstract}
In the cosmological settings, Quantum Gravity effects are typically understood to be limited towards very early phase of the universe, namely in the pre-inflationary era, with limited signatures spilling over into the succeeding inflationary era which gradually fade away as the universe transits into the subsequent radiation and matter driven expansion.  In general scenarios as well, the imprints of possible quantum character of gravitons are typically so feeble that they are expected to remain buried under overwhelming noise from other stronger processes. In this work, we demonstrate that quantized gravitational perturbations cause strong observable effects in cosmological settings post the Last Scattering Surface (LSS) more prominently than any other classical or quantum processes. This counter-intuitive effect is facilitated  by the fact that the quantum correlators of the gravitons {\it grow divergently large in the matter dominated era} unlike any other background fields leading to an abrupt rise in the 
processes mediated by correlators of quantum gravitons, which otherwise is way too feeble.  The transitions between spherical harmonics states of a  hydrogen atom guided by gravitons in a particular phase of the late time era of the universe, provides an example of such a process.  Thus the newly formed atoms post LSS are expected to get excited from the vacuum fluctuations of gravitons quite efficiently in this era and also get de-excited to convert the vacuum fluctuations to an additional channel of energy influx into the evolution history. Such unavoidable processes provide an avenue of quantum graviton mediated process becoming important in the late time era, which has many interesting implications.
\end{abstract} 

\maketitle
\section{Introduction}
The development of a concrete theory of Quantum Gravity has remained elusive, partly due to the lack of any experimental, primarily due to the fact that for available energy scales, the quantum gravity processes are typically  so heavily suppressed that occurrence of any such phenomena is phenomenally rare. There are always other processes with much larger amplitudes, that  over-dominate the contributions from any quantum gravity channels \cite{Kiefer:2005uk, Boughn:2006st, scully_zubairy_1997,milonni1994quantum, loudon2000quantum, Martin-Martinez:2012ysv}.
Apart from having a direct access to the regions of spacetime curvature singularities, where quantum gravity is understood to be directly present in the play, another route is to look for the processes which are assisted by quantum gravity particles - the quantum gravitons. Irrespective of the exact details of any full quantum gravity theory, their linear order behaviour, i.e. simple properties of gravitons, are supposed to be in agreement. There have been proposals to unravel the quantum aspects of graviton e.g. the transitions caused by graviton correlators \cite{Guerreiro:2019vbq,Parikh:2020nrd,Anastopoulos:2020cdp,Parikh:2020kfh,Parikh:2020fhy} or development of entanglement between quantum systems \cite{Marletto:2017kzi,Bose:2017nin,krisnanda2020observable,Belenchia:2018szb,Carney:2018ofe,Kanno:2021gpt}. Yet,  the strength of such processes are typically so feeble that any other source/mediator (e.g. electromagnetically induced)  will cause the transitions/entanglements much effectively and much faster 
before quantum gravitational perturbations would be able to do that \cite{Rothman:2006fp,Jones:2015uda}.  Thus, in the standard settings, occurrence of any such processes can not be attributed to quantum gravitons with definiteness.

In cosmological settings, the early universe is a regime where quantum gravity is expected to implant some imprints very early on, into the evolutionary story of the universe. Study of inflationary modes is a well understood approach of unravelling some of the aspects of quantum gravity at work, through the study of the so-called trans-Planckian effects \cite{ Martin:2000xs, Damour:1995pd,Tsujikawa:2003vr}. The mode evolution of the quantum fields during the pre-inflationary era are expected to be sensitive to the dynamics of that era and there may be observables which preserve some of the qualitative imprints \cite{Bojowald:2002nz}. Still, as the universe grows and enters later epochs of radiation and matter dominance, the quantum features are understood to be diminished and the universe takes up a classical description \cite{Kiefer:2008ku,Brandenberger:1990bx}. The perturbations from the standard classical evolution lead to some observable effects \cite{CMBPolStudyTeam:2008rgp} in the cosmic microwave 
background radiation (CMBR) -- our primary messenger from the earliest epochs of the evolution, but it is widely believed that all quantum effects must have died by the time of CMBR formation and the subsequent evolution is -- by and large -- a classical phenomenon. Thus search of a quantum gravity needle in the classical haystack becomes a formidable task, if not impossible.

In this work, we argue that counter to standard intuition, late time era in cosmology also provides an avenue where some quantum gravity mediated processes get suddenly {\it hugely revived} and and dominating so much so that {\it certain physical processes are totally dictated by quantum gravity perturbations than by any other field}.  The basic reason for this late time quantum gravity revival is the fact the graviton correlators become divergently large in a strong matter dominated era, very similar to the early inflationary era. Therefore, the processes which are sensitive to graviton contributions pick up the {\it  late time revival of graviton correlators}. For instance, the hydrogen atoms formed in the late matter dominated era  -- post the Last Scattering Surface (LSS) -- can undergo transitions between its internal spherical harmonics states due to interaction with quantum gravitons with new selection rules, a process which sharply rises in deep matter dominated era.  We show that the photons mediated 
transitions, which typically is the most efficient channel,  become secondary mode of transition for a ``brief period'' of time in the late time era. Thus, such transitions may leave late time imprints to the CMBR sky and provide a direct way of observing quantum gravity effects present in the late time era. The processes we consider are purely  quantum graviton mediated, without any competitor i.e. there is no appreciable contribution from any classical (or even quantum) field which will cause transitions as strongly as the gravitons do and subsequently set up a new equilibrium configuration in the late matter domination era of cosmic expansion. 

Further, what an excited atom would emit (photons, gravitons or any other particle)  during the de-excitation transition depends upon the coupling to the respective quantum fields and the strength in their correlators. If we quantize both the electromagnetic field as well as the graviton field, their individual interaction with atoms are primarily through dipole and quadrupole couplings respectively.  In a  standard flat space an atom is typically coupled most strongly with the EM field (the gravity coupling is some $10^{45}$ orders small).  Thus the probability of a photon emission is much larger than that of a graviton emission (roughly by $10^{45} : 1$ ratio). {\it What we show in this paper is that in the strong matter dominated era phase of the universe, despite this weak coupling ratio, the graviton correlator grows divergently large and the effective probability swings in the favour of  graviton emission.} Thus in this era an atom will primarily make transitions through this channel. Such a system 
will lead to a burst of gravitons during de-excitation  in the matter dominated era, analogous to the spontaneous emission, which provides an additional energy content into the evolution profile. All such effects are expected to leave an visible imprint  of the revival of quantum gravity at the late time era on the CMB sky as well as the evolution profile of universe, some of them we discuss at the end.

\section{Gravitational Wave as massless scalar field in FRW universe}
To study the propagation of gravitational waves in flat spacetime, linearized metric perturbations are studied on the background of a Minkowski metric.  These gravitational perturbations are known to satisfy the wave equation in the flat spacetime which other massless fields such as the  electromagnetic fields also satisfy \cite{Kiefer:2005uk,Parker:2009uva}. However, if we take the background spacetime to be a conformally flat one, such as the Friedmann spacetime, this equivalence breaks. The electromagnetic field, which in  3+1 dimensions is a  conformal field, keeps satisfying the same wave equation it would have in the flat spacetime, whereas the gauge invariant gravitational perturbations start satisfying a wave equation what a non conformal minimally coupled massless scalar field would \cite{Weinberg:2008zzc,Caprini:2018mtu}
\bea
\ddot{h}_{ij}(k) + 3 \frac{\dot{a}}{a}\dot{h}_{ij}(k)  + k^2 h_{ij}(k)=0. \label{GWEoM}
\eea
Due to the breaking of the conformal invariance, massless scalar field and gravitational fields undergo one of the most interesting prediction about quantum fields on curved spacetime -{\it particle creation} \cite{Parker:1968mv,Parker:1969au,Birrell:1982ix,Mottola:1984ar}. No particle creation takes place for electromagnetic field on the other hand \cite{Parker:2009uva}.
Apart from particle creation, gauge invariant tensor perturbations (gravitational waves) also share another aspect with the scalar fields in Friedmann spacetimes -- their non trivial  quantum correlation structure, which is the source of the late time effects we consider in this paper and is what we  discuss next.
\subsection{Correlator structure of minimal massless fields in Friedmann universe}
The action of a minimally coupled massless scalar field in a power law Friedmann universe with metric, $g_{\alpha\beta}= a^2(\eta) \eta_{\alpha\beta},$ is given by
\begin{eqnarray}
S&\equiv& -\frac{1}{2}\int d^{4}x \sqrt{-g}g^{\mu \nu}\partial_{\mu}\phi\partial_{\nu}\phi \nonumber\\
&=&-\frac{1}{2}\int d^{4}x \, a^4\big(a^{-2}\,\eta^{\mu \nu}\partial_{\mu}\phi \,\partial_{\nu} \phi\big)\, .\label{ActionMassless}
\end{eqnarray}
For a power law Friedmann universe described by $a(\eta) = (H\eta)^{-q}$, under the conformal transformation, $\phi(x) = (H\eta)^{-1+q}\, \psi(x)$, the action can be re-written as that of a {\it massive scalar field  $\psi(x)$ in the de Sitter universe with scale factor} $\tilde{a}(\eta) = -(H\eta)^{-1}$,
\begin{eqnarray}
S&\equiv& -\frac{1}{2}\int d^{4}x \sqrt{-g}g^{\mu \nu}[\partial_{\mu}\psi\partial_{\nu}\psi -m_{eff}^2 \psi^2 ] \nonumber\\
&=& -\frac{1}{2}\int d^{4}x \,\tilde{a}^4\big(\tilde{a}^{-2}\,\eta^{\mu \nu}\partial_{\mu}\psi \,\partial_{\nu} \psi - m_{eff}^2 \psi^2 \big)\, ,
\end{eqnarray}
having mass defined by $m^2_{eff} = H^2(1 - q)(2 + q)$ \cite{Lochan:2018pzs}. Various aspects of this duality have been studied in \cite{Lochan:2018pzs, Dhanuka:2020yxp,Lochan:2022dht}.  {\it Curiously a massless field in the matter-dominated era ($q=-2$) maps to a massless field in the de Sitter space.} 

Owing to the conformal connection between the theories, the Wightmann function in the power law Friedmann universe is related conformally to one in the de Sitter spacetime  i.e., $G_{FRW}(x,y;m^2=0) = (H\eta)^{q-1}(H\eta')^{q-1}G_{dS}(x,y;m^2_{eff}=H^2(1-q)(2+q))$.  For the matter dominated era i.e., $q=-2$,  one has $G_{\text{Matter}}(x,y;m^2=0) = (H^2\eta \eta')^{-3}G_{dS}(x,y;m^2_{eff}=0)$, which is well known to support large correlations -a primary reason for the scale invariant power spectrum \cite{Lochan:2018pzs}.  Interestingly, being massless, the gauge invariant tensor perturbations do follow the same action, equation of motion and hence the same correlation, upto an elementary tensorial fucntion $\Big(\delta_{ip}\delta_{jk} + \delta_{ik}\delta_{jp} - \delta_{ij}\delta_{pk} \Big)
$ and hence have the same conformal structure in the Wightmann functions between the matter dominated era and the de Sitter spacetime, which we employ in analysing the graviton assisted transitions in the hydrogen atom (see Appendix for details).
\section{Hydrogen atom in cosmology}
In the baryogenesis discussion in cosmology, it is well understood that neutral hydrogen formation happened when CMBR photons did not possess enough energy to ionize the  hydrogen atoms, a time around a redshift of $z\sim 1100$ (LSS). The newly formed hydrogen atoms at the LSS could now play the role of a quantum system interacting with background quantum fields available to couple with, and its transitions be used as a probe of the quantum correlations of the background field.
\begin{figure}[h]
\includegraphics[scale=0.45]{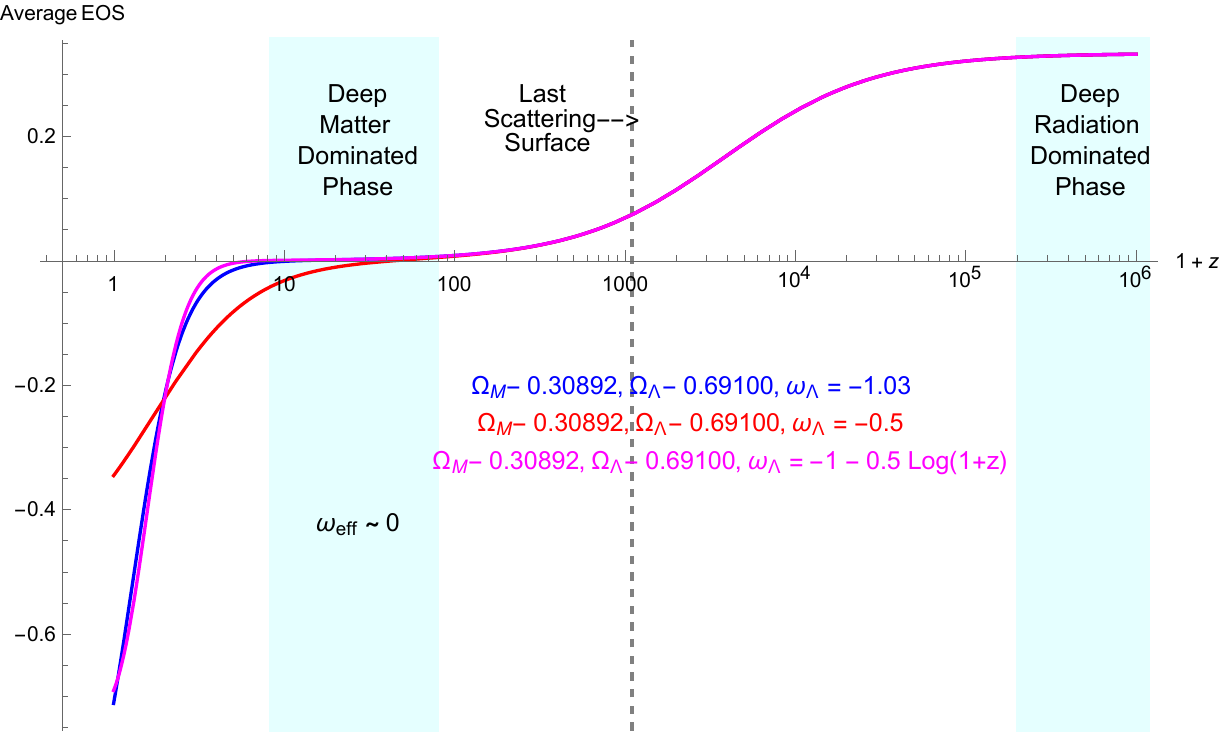} 
\caption{The average Equation of State (EOS) during the evolution of the universe across various cosmological models grazes $w_{eff} \sim 0$ in the window $20<z<100$.}\label{Fig1}
\end{figure} 

After their formation, neutral hydrogen atoms free stream in progressively more matter dominated era  before structure formation begins $(z<20)$ and ultimately dark energy kicks in \cite{White:1994bn}. Somewhere in between these atoms stream by in an era where {\it effective} equation of state is completely matter-like $w_{eff} \sim 0$ (see fig.(\ref{Fig1}) giving the universe an effective description of the kind $ a(\eta) \rightarrow (H\eta)^{2-\delta}$ with $\delta \rightarrow 0$ \footnote{We can approximate the scale factor with a constant power law universe for a duration if $\dot{\delta} /\delta \sim \dot{\cal H} /{\cal H} \ll {\cal H}$, near the era when $w_{eff}$ grazes zero, where ${\cal H}$ is the Hubble parameter of expansion.}. Consequently it leads to a development of a tiny mass to the dual de-Sitter field $m^2_{eff} \sim  3 H^2\delta\rightarrow 0 $ \footnote{ We provide analysis for a late time vacuum state in the matter dominated era but the divergent structure remains true for all physically 
well behaved states as correlation in a general state contains the correlator structure of vacuum as well \cite{Birrell:1982ix, Lochan:2022dht, Lochan:2014xja}}. This vanishingly small mass accords a large correlation to the fields in de Sitter spacetime as the Wightmann function for this vanishing mass is given by \cite{Allen:1987tz,Lochan:2018pzs}
\begin{equation}
G_{dS}(x,x') \propto \Big(\frac{H^2}{16\pi^2}\Big)\Big(\frac{2}{\delta}+ \frac{4}{y} -4 - 2ln(y) + 4ln2 + O(\delta)\Big) \, , \label{WFDS}
\end{equation} 
where $y$ is the de Sitter invariant distance between $x$ and $x'$.

Thus both in the de Sitter and matter dominated era, the Wightman function diverges for the non-conformal massless scalar fields and would do similarly so for gravitons. Even for an era very near to the matter domination, the correlators will be exceedingly large as
$G_{\text{Near-Matter}}(x,x';m^2=0) 
=(H^2\eta_{1}\eta_{2})^{-3 + \delta} G_{dS}(x,x';m^2_{eff} \sim  3 H^2\delta\rightarrow 0)$. The hydrogen atoms, in interaction with the quantum gravitational perturbations take up the structure of a derivatively coupled Unruh DeWitt detector (UDD) as we shall see below.  Unlike the simple coupling, the structure of derivative coupling has a peculiar character in Friedmann universes -- the derivative coupling regularizes the divergences appearing in the Wightman function for the de sitter era, {\it but still maintains the divergences in the matter dominated era }\cite{dhanuka2022unruh}. Divergences appearing in the correlators of a quantum field are known to cause rapid transition in UDD \cite{Stargen:2021vtg} even under mild conditions. Now that gravitons have divergent correlator in a window of the matter dominated era, they also cause rapid transitions in the hydrogen atom even with a weak curvature in the window $z \sim 20-100$ when the universe is undergoing almost complete matter driven expansion, thus 
leaving robust imprints of graviton mediated transitions. We will also show the complete dominance of quantum graviton driven transition over other channels of "noise".

\subsection{Coupling of Hydrogen atoms with gravitational waves}
Since post LSS, the hydrogen atoms are expected to move non-relativisitically \cite{Kolb:1990vq},  we can employ the non-relativistic limit of the Dirac equation resulting in the Schr\"{o}dinger equation for the electron \cite{Parker:1980kw,Parker:1980hlc,Parker:1982nk,bessis1984,dhanuka2022unruh}
\begin{equation}
\left(i\frac{\partial}{\partial t} -m_e\right)\psi = \left(-\frac{1}{2m_e}\nabla^2- \frac{e^2}{r} + \frac{1}{2}m_e R^{FNC}_{0p0q}x^{p}x^{q}\right)\psi,
\end{equation}
upto leading order in  the Fermi Normal Co-ordinates (FNCs) $x^l$, where $R_{0l0m}$ are the Riemann tensor components expressing the curvature induced corrections to the flat spacetime Schr\"{o}dinger equation with the central Coulomb potential of the nucleus \cite{Parker:1980kw,dhanuka2022unruh}. The Riemann tensor components  in FNCs are related to those in any arbitrary coordinate system, by the relation
$R^{FNC}_{abcd} = R^{arbitrary}_{\mu\nu\gamma\delta}\vec{e}^{\mu}_{a}\vec{e}^{\nu}_{b}\vec{e}^{\gamma}_{c}\vec{e}^{\delta}_{d} \,
$
where $\vec{e}^{\mu}_{a}$ are a set of orthonormal basis parallel transported along the central timelike geodesic. The $\vec{e}^{\mu}_{0}$ is the tangent vector field along the central geodesic  the hydrogen atom follows \cite{Poisson:2009pwt,dhanuka2022unruh}. \\
Treating curvature induced terms as perturbations, one employs time dependent perturbation theory to obtain rate of transition of the hydrogen atom  \cite{Juarez-Aubry:2014jba,Juarez-Aubry:2018ofz}. We analyse the case when the background spacetime is a perturbed Friedmann universe and the (tensor) perturbations are quantized \cite{Parker:2009uva}.

\section{Transitions in perturbed FRW spacetimes}
A Friedmann universe supporting tensor perturbations can be expressed in conformal time co-ordinates as
\begin{equation}
ds^2 = a^2(\eta)(\eta_{\mu\nu} + h_{\mu\nu})dx^{\mu} dx^{\nu} \, .
\end{equation}
Using the orthonormal basis vector fields connecting the conformal co-ordinates to FNCs, we can convert the Riemann tensor from conformal coordinates to the Fermi normal coordinates (FNCs) as (See Appendix)
\begin{equation}
R^{FNC}_{0p0q} = \frac{1}{a^4}\Big(-\delta_{pq}(aa'' - a'^2) -\frac{aa'}{2}h_{pq}^{'} -\frac{a^2}{2} h_{pq}^{''}\Big)  + O(h^2).
\end{equation}
Using the above expression for the relevant Riemann components, the interaction Hamiltonian in the co-moving frame is obtained as
$H_{I} = m_e H_{pq}x^{p}x^{q}/2$
with $H_{pq} = \Big(-\delta_{pq}\frac{\ddot{a}}{a} -\frac{\dot{a}}{a}\dot{h}_{pq} -\frac{1}{2} \ddot{h}_{pq}\Big)$. Here we can see that the interaction term takes form of a {\it generalized UDD interaction with derivative coupling} where the part $m_ex^{p}x^{q}/2$ belongs to the quantum system and the part $ H_{pq}$  to the background field.
Similar to the selection rules for electromagnetic transitions, the  quantized operator $\hat{H}_{I} = m_e \hat{H}_{pq}\hat{x}^{p}\hat{x}^{q}/2$ will offer selection rules for the graviton mediated transitions, see Table \ref{SelectionRulesI} (see Appendix for details).
\begin{table}[h]
\centering
\begin{tabular}{|cc|cc|cc|}
\hline
\multicolumn{2}{|c|}{$x^2, y^2, xy, yx$}      & \multicolumn{2}{c|}{$xz, yz$}              & \multicolumn{2}{c|}{$z^2$}                   \\ \hline
\multicolumn{1}{|c|}{$\Delta l$} & $\Delta m$ & \multicolumn{1}{c|}{$\Delta l$} & $\Delta m$ & \multicolumn{1}{c|}{$\Delta l$} & $\Delta m$ \\ \hline
\multicolumn{1}{|c|}{-2,0,2}     & -2,0,2     & \multicolumn{1}{c|}{-2,0,2}     & -1,1       & \multicolumn{1}{c|}{-2,0,2}     & 0          \\ \hline
\end{tabular}
\caption{Selection rules for transition elements $<n',l',m'|x^{i}x^{p}|n,l,m>$.} \label{SelectionRulesI}
\end{table}

Taking into account the selection rules, let us now consider an allowed transition between different spherical harmonics $\{nlm\}\to \{n'l'm'\}$. For these transitions, the unperturbed curvature induced term proportional to $\delta_{pq}$ in $H_{pq}$ drops out for $l\neq l', m\neq m'$ and the vacuum expectation of the product of gravitational perturbations appearing in the transition probability is 
\begin{eqnarray}
\lim_{\vec{y}_1 \to \vec{y}_2}\bra{0} \hat{H}_{ij}(\vec{y}_{1},\eta_{1})\hat{H}_{pk}(\vec{y}_{2},\eta_{2})\ket{0}  = \lim_{\vec{y}_1 \to \vec{y}_2}\frac{1}{a_{1}^{4}}\Big(\frac{a_{1}a_{1}^{'}}{2}\frac{\partial}{\partial\eta_{1}} -\frac{a_{1}^{2}}{2} \nabla^{2}_{\vec{y}_{1}}\Big)\frac{1}{a_{2}^{4}}\Big(\frac{a_{2}a_{2}^{'}}{2}\frac{\partial}{\partial\eta_{2}} -\frac{a_{2}^2}{2} \nabla^{2}_{\vec{y}_{2}}\Big)\underbrace{\bra{0} \hat{h}_{ij}(\vec{y}_{1},\eta_1)\hat{h}_{pk}(\vec{y}_{2},\eta_2)\ket{0}}_{G_{ijpk}}. 
\end{eqnarray}
Thus, we see that the graviton correlator $G_{ijpk}$ appears naturally in the transition probability employing which (see Appendix), we get
 \begin{widetext}
\begin{multline}\label{TransitionProb}
P_{\psi_{nlm}\to \psi_{n'l'm'}} = \frac{m_e^2}{4}\bra {\psi_{n'l'm'}}\hat{x}^{i}\hat{x}^{j}\ket{\psi_{nlm}}^{*}\bra {\psi_{n'l'm'}}\hat{x}^{p}\hat{x}^{k}\ket{\psi_{nlm}}   \lim_{\vec{y}_1 \to \vec{y}_2}\int^{\eta_{f}}_{\eta_{i}}d\eta_{1}\int^{\eta_{f}}_{\eta_{i}}d\eta_{2}e^{-i\Omega(t(\eta_{1})-t(\eta_{2}))}\\ \frac{1}{a_{1}^{3}}\Big(\frac{a_{1}a_{1}^{'}}{2}\frac{\partial}{\partial\eta_{1}} \Big)\frac{1}{a_{2}^{3}}\Big(\frac{a_{2}a_{2}^{'}}{2}\frac{\partial}{\partial\eta_{2}} \Big)\Big(\delta_{ip}\delta_{jk} + \delta_{ik}\delta_{jp} - \delta_{ij}\delta_{pk} \Big)(H^2\eta_{1}\eta_{2})^{-3 + \delta}\frac{H^2}{32\pi^2 \delta} +  \text{ Subdominant terms}.
\end{multline}
\end{widetext}
One can further define a rate of transition by going to $\tilde{\eta},\eta_- \sim \eta \pm \eta'$ basis as (see Supplemntary naterial) 
\begin{equation}
\Gamma_g =dP_{\psi_{nlm}\to \psi_{n'l'm'}} /d \tilde{\eta} \sim \frac{m_e^2  Q^2 \Omega^5}{\delta}(H/\Omega)^{-\frac{q}{1-q}},
\end{equation}
where  $Q^2 = |\bra {\psi_{n'l'm'}}\hat{x}^{i}\hat{x}^{j}\ket{\psi_{nlm}}|^2 \sim a_0^4$ represents the quadrupole moment elements and $a_0$ the Bohr radius.
One can see that in the era of $\omega_{eff}\sim 0$ i.e., $\delta \sim 0$ arbitrary small, we have an abrupt rise in the rate of transitions mediated by quantum $h_{\mu \nu}$. Note that the cause of the abrupt rise {\it  is solely due to conformal perseverance of the de-Sitter divergent correlators under the derivative coupling}. In the de Sitter spacetime itself such a divergence would not have survived as being spacetime independent --Eq.(\ref{WFDS}) -- the derivative actions of the coupling annihilate the divergent term. Further, in other era e.g.- the radiation dominated era, the graviton correlator remains finite and does not lead to any enhanced transitions as the dual theory in de Sitter is massive, devoid of any divergence \cite{Lochan:2018pzs, dhanuka2022unruh}. {\it It is in the deep matter dominated era only where the dominance of quantum graviton mediated transition is manifest}. 

In comparison, the other long range interaction over the cosmological scale is the electromagnetic 
field, which in 3+1-dimensions is a conformal field. Thus in conformal time co-ordinate, its correlators behave exactly like in flat spacetime and do not blow up ever. An elementary analysis akin to that in the flat space shows that the transition probability and rate can neatly be expressed as (see Appendix)

\begin{eqnarray}\label{TransitionProbE}
P_{0 \to \Omega}&=& d^2 \int_{\eta_{i}}^{\eta_{f}} \int_{\eta_{i}}^{\eta_{f}} d\eta_{1} d\eta_{2}e^{-i\Omega(t(\eta_{1})-t(\eta_{2}))} \frac{1}{\pi^2 (\Delta \eta - i\epsilon)^4}\,\nonumber\\
\Gamma_e &\sim& d^2 \Omega^3\left(\frac{H}{\Omega}\right)^{\frac{-3q}{1-q}} \,
\end{eqnarray}
with $d^2=|\bra{\psi_{n'l'm'}}\hat{d}^{z}(0)\ket{\psi_{nlm}}|^2  \sim e^2 a_0^2$ being the dipole moment transition element of a hydrogen atom where $e$ is the electron charge. In the flat space case $a(\eta) =1; \eta =t$ gives the familiar result \cite{Kiefer:2005uk, Boughn:2006st} that for hydrogen atom involving states with $\Omega \sim 10 eV$ the rate of graviton induced and photon induced transitions (restoring the factors $G,\hbar,c,\epsilon_0$ in terms of fine structure constant $\alpha$ and Planck mass $m_{Pl}$) have the hierarchical structure  $(\Gamma_g/\Gamma_e)_{\text{flat}} \sim   \frac{m^2 Q^2 \Omega ^2}{d^2 } \sim   \alpha^2 (m_e/m_{Pl})^2 \sim 10^{-45}.$ On the other hand, for a near-matter driven universe with $a(\eta) = (H \eta)^{2-\delta}$,  the graviton to electromagnetic transition rate ratio from Eq. (\ref{TransitionProb},\ref{TransitionProbE}), takes the form
\begin{equation}
\lim_{\delta \rightarrow 0}\Big(\frac{\Gamma_{g}}{\Gamma_{e}}\Big)_{\delta} \propto  \lim_{\delta \rightarrow 0} \Big(\frac{\Gamma_{g}}{\Gamma_{e}}\Big)_{\text{flat}} \frac{1}{\delta} \Big(\frac{H}{\Omega}\Big)^{\frac{2 q}{1-q}}.
\end{equation}


One sees that as $\delta \rightarrow 0$ (which is equivalently $q\rightarrow  -2$) the factor  $ (H/\Omega)^{-4/3}\times 1/\delta$ swings the ratio drastically in the favor of gravitons. For transitions of $\sim 10 eV \Rightarrow \Omega \sim 10^{16}s^{-1}$  close to $z\sim 30$ where the parameter, ~ $H \sim H_0[\Omega_{M,0}(1+z)^3/(\Omega_{\Lambda, 0} +\Omega_{M,0})]^{1/2} \sim10^{-16}s^{-1}$, for $\delta \rightarrow 10^{-3}$ and smaller, this factor provides an astronomical enhancement  $\sim 10^{47}$.  Hence, the deep matter dominated era $(\delta \ll 1)$ provides a singular epoch of quantum graviton dominance over other forces. As the universe exists this phase the graviton contribution quickly fizzles out.

\section{Observational Implications}
We have seen that graviton mediated transition rate of the hydrogen atom shoots up abruptly when the universe undergoes a near perfect matter dominated phase $\delta \rightarrow 0$. This will lead to various interesting physical phenomena which we briefly discuss.

\begin{itemize}

\item The rapid transitions are primarily mediated by vacuum graviton correlations\footnote{If we take the classical gravitons instead, in the transition amplitude the derivatives of the classical product $h_{ij}h_{lm}$ will appear which will be abysmally small, unlike the quantum correlator}. The upward transitions will happen primarily through vacuum fluctuations but the de-excitations will cause an extra graviton emission, akin to spontaneous emission, causing a burst of gravitons which will further lead to visible effects of CMB polarizations \cite{CMBPolStudyTeam:2008rgp}. The number of such gravitons will be proportional to the number of bound atoms  (any bound atom will also undergo similar transitions many times during the era $\omega_{eff} \sim 0$) and thus provides a lower bound on polarization. 

\item Further, the infusion of gravitons leads to conversion of energy from vacuum to real particles, contributing as an additional source of energy in late time era, pertinent to the recent debate of apparent upshift of the Hubble parameter visualized from the low red shift surveys \cite{DiValentino:2021izs}, note that $H^2 \sim \sum \rho$.  In fact, since gravity is expected to couple to dark matter also uniformly, if the dark matter atoms also have a similar bound state structure, they will also contribute to such energy extraction from the graviton vacuum. 

\item It is easy to see from  Eq.~\ref{TransitionProb} that the leading order divergent term is symmetric under $\Omega\rightarrow -\Omega$. In terms of thermality, this condition is achieved in Unruh deWitt Detectors if they are put in {\it infinite temperature bath} (also  see \cite{dhanuka2022unruh}). Therefore the hydrogen atoms will undergo rapid excitations and de-excitations to disrupt the standard thermal distribution they would have established w.r.t. the background CMBR (which will be of the order of $60-300 K$ around the window where $w_{eff} \sim 0$ [see fig.(\ref{Fig1})]). Under the action of the graviton mediated rapid transitions the hydrogen atoms will renegotiate a new equilibrium condition $n_e/n_g \sim\Gamma_{\uparrow}/\Gamma_{\downarrow} \sim 1$. As the universe grows out of the matter dominated phase and the graviton driven transitions gradually fade away \cite{dhanuka2022unruh}, such excited atoms will start emitting through other channels as well and may leave optical imprints \cite{Bowman:2018yin}.

\end{itemize}

Thus {\it  the late time strong matter dominated epoch of the universe provides a unique window to have a glimpse of quantum gravity processes and may potentially be used to test and constrain quantum gravity models against available observable channel.} A more detailed analysis of many such potential effects will be pursued elsewhere.
\section*{Acknowledgments}
\noindent AD would like to acknowledge the financial support from University Grants Commission, Government of India, in the form of Junior Research Fellowship (UGC-CSIR JRF/Dec-2016/510944). Research of KL is partially supported by the Department of Science and Technology (DST) of the Government of India through a research grant under INSPIRE Faculty Award (DST/INSPIRE/04/2016/000571) and a MATRICS research grant no.
MTR/2022/000900 from the SERB. Authors also acknowledge Wolfram Mathematica which has been used to plot some of the figures. 

\section*{Appendix}
\setcounter{section}{0}
\section{Action formulation and correlator structure for tensor perturbation}
\noindent The action for tensor perturbations to the background FRW dynamics of the Universe, $h_{ij}$, is given by \cite{Brandenberger:1990bx}
\begin{equation}
    S = \frac{1}{8 } \int d\eta\ d^3\vec{y}\ a^2 \Big((h_{ij}^{'})^2 - (\nabla h_{ij})^2\Big) \, ,
\end{equation}
where $h_{ij}\delta^{ij} = 0$ and $\partial^{i}h_{ij} = \partial^{j}h_{ij} = 0.$ The above action for each component of the perturbations is just the same as that of a massless scalar field in an FRW background and hence they satisfy the same equation of motion. However, the transverse traceless conditions on the tensor perturbations leave only two independent components and hence the tensor perturbations are dynamically equivalent to two massless scalar fields in FRW backgrounds. \\
The quantized tensor perturbations can be expanded as 
\begin{equation}\label{2}
\hat{h}_{ij}(\vec{y},\eta) = \sum_{\lambda = +,\times}\int d^3\vec{q}\ e_{ij}(\hat{q}, \lambda) \Big(e^{i\vec{q}.\vec{y}}h_{q}(\eta)\hat{b}_{\vec{q},\lambda} + e^{-i\vec{q}.\vec{y}}h^{*}_{q}(\eta)\hat{b}^{\dagger}_{\vec{q},\lambda}\Big) \, ,
\end{equation}
where $\lambda = +, \times$ refer to two polarization states and $\hat{b}_{\vec{q},\lambda}$ and $\hat{b}^{\dagger}_{\vec{q},\lambda}$ are the annihilation and creation operators for state with wave-vector, $\vec{q}$, and polarization, $\lambda$.  
 The $e_{ij}(\vec{q},\lambda)'s$ satisfy \cite{Weinberg:2008zzc}
\begin{multline}\label{3}
\sum_{\lambda = +,\times}e_{ij}(\hat{q},\lambda)e_{kl}(\hat{q},\lambda) = \delta_{ik}\delta_{jl} + \delta_{il}\delta_{jk} - \delta_{ij}\delta_{kl} + \delta_{ij}\hat{q}_{k}\hat{q}_{l} + \delta_{kl}\hat{q}_{i}\hat{q}_{j} \\ -\delta_{ik}\hat{q}_{j}\hat{q}_{l}-\delta_{il}\hat{q}_{j}\hat{q}_{k}-\delta_{jk}\hat{q}_{i}\hat{q}_{l}-\delta_{jl}\hat{q}_{i}\hat{q}_{k} + \hat{q}_{i}\hat{q}_{j}\hat{q}_{k}\hat{q}_{l} \, .
\end{multline}
The time dependent part, $h_{q}(\eta)$, satisfies  \begin{equation}\label{9}
h_{q}^{''}(\eta) + 2\frac{a'}{a}h_{q}^{'}(\eta) + q^2 h_{q}(\eta) =0 \, ,
\end{equation}
making the time evolution independent of the direction of wave-vector and polarization state.

\noindent Employing equation (\ref{2}) and the commutation relations between creation and annihilation operators, we get
\begin{eqnarray}\label{6}
\bra{0} \hat{h}_{ij}(\vec{y}_{1},\eta_1)\hat{h}_{pk}(\vec{y}_{2},\eta_2)\ket{0} &=& \sum_{\lambda = +,\times}\int d^3\vec{q}\ e_{ij}(\hat{q}, \lambda) e_{pk}(\hat{q}, \lambda)e^{i\vec{q}.(\vec{y}_{1}-\vec{y}_{2})}h_{q}(\eta_{1})h^{*}_{q}(\eta_{2}) \nonumber \\
&=& \int d^3\vec{q}\ \Big(\sum_{\lambda = +,\times}e_{ij}(\hat{q}, \lambda) e_{pk}(\hat{q}, \lambda)\Big)e^{i\vec{q}.(\vec{y}_{1}-\vec{y}_{2})}h_{q}(\eta_{1})h^{*}_{q}(\eta_{2}).
\end{eqnarray}
Using (\ref{3}) and (\ref{6}), the correlator between tensor perturbations is seen to be given by
\begin{eqnarray}\label{8}
\bra{0} \hat{h}_{ij}(\vec{y}_{1},\eta_1)\hat{h}_{pk}(\vec{y}_{2},\eta_2)\ket{0} &= & \Big(\delta_{ip}\delta_{jk} + \delta_{ik}\delta_{jp} - \delta_{ij}\delta_{pk} + \delta_{ij}\frac{\partial_{\vec{y}_{1p}}\partial_{\vec{y}_{1k}}}{\nabla_{\vec{y}_1}^{2}} + \delta_{pk}\frac{\partial_{\vec{y}_{1i}}\partial_{\vec{y}_{1j}}}{\nabla_{\vec{y}_1}^{2}}  -\delta_{ip}\frac{\partial_{\vec{y}_{1j}}\partial_{\vec{y}_{1k}}}{\nabla_{\vec{y}_1}^{2}} \nonumber \\ &&- \delta_{ik}\frac{\partial_{\vec{y}_{1j}}\partial_{\vec{y}_{1p}}}{\nabla_{\vec{y}_1}^{2}}-\delta_{jp}\frac{\partial_{\vec{y}_{1i}}\partial_{\vec{y}_{1k}}}{\nabla_{\vec{y}_1}^{2}}-\delta_{jk}\frac{\partial_{\vec{y}_{1i}}\partial_{\vec{y}_{1p}}}{\nabla_{\vec{y}_1}^{2}} + \frac{\partial_{\vec{y}_{1i}}\partial_{\vec{y}_{1j}}\partial_{\vec{y}_{1p}}\partial_{\vec{y}_{1k}}}{\nabla_{\vec{y}_1}^{2}\nabla_{\vec{y}_1}^{2}}\Big)\nonumber \\ && \hspace{2cm} \underbrace{\int d^3\vec{q} e^{i\vec{q}.(\vec{y}_{1}-\vec{y}_{2})}h_{q}(\eta_{1})h^{*}_{q}(\eta_{2})}_{G_{\text{FRW}}(x(\eta_{1}),x(\eta_{2}))} \, .
\end{eqnarray} 
where $G_{FRW}$ is the Wightman function for massless scalar fields in FRW spacetimes. 
\section{Transitions in perturbed FRW spacetimes}

\noindent A Friedmann universe supporting tensor perturbations can be expressed in conformal time co-ordinates as
\begin{equation}
ds^2 = a^2(\eta)(\eta_{\mu\nu} + h_{\mu\nu})dx^{\mu} dx^{\nu} \, .
\end{equation}
Since the perturbations interaction is written in the co-moving frame of hydrogen atoms one needs to transport the  curvature tensor in these co-ordinates to the FNCs.

It can be shown (see \cite{Dai:2015rda,dhanuka2022unruh}) that the set of parallel transported orthonormal basis upto first order in $h$ is given by 
\begin{eqnarray}
\vec{e}^{\mu}_0 &=& \frac{1}{a}(1 ,0) \, ,\\ \vec{e}^{\mu}_i &=& \frac{1}{a}(0,\delta^{j}_{i}-\frac{h^{j}_{i}}{2} ) \, .
\end{eqnarray}
Using these orthonormal basis vector fields, we can convert the Reimann tensor from conformal coordinate system to the Fermi normal coordinates (FNCs) by the relation 
\begin{equation}
R^{FNC}_{abcd} = R^{Con}_{\mu\nu\gamma\delta}\vec{e}^{\mu}_{a}\vec{e}^{\nu}_{b}\vec{e}^{\gamma}_{c}\vec{e}^{\delta}_{d} \, .
\end{equation}
In particular, we have 
\begin{eqnarray}
R^{FNC}_{0p0q} = \frac{1}{a^4}R^{Con}_{0k0p}\big(\delta^{k}_{p}-\frac{h^{k}_{p}}{2}\big)\big(\delta^{p}_{q}-\frac{h^{p}_{q}}{2}\big) \, .
\end{eqnarray}
The Riemann tensor components relevant for our purposes, in conformal coordinates, are given by 
\begin{equation}
R^{Con}_{0p0q} = -\delta_{pq}(aa'' - a'^2) - (aa''-a'^2)h_{pq}-\frac{aa'}{2}h_{pq}^{'} -\frac{a^2}{2} h_{pq}^{''} \, .
\end{equation}
Using this, we see that 
\begin{eqnarray}
R^{FNC}_{0p0q} 
& = & \frac{1}{a^4}\Big(-\delta_{pq}(aa'' - a'^2) -\frac{aa'}{2}h_{pq}^{'} -\frac{a^2}{2} h_{pq}^{''}\Big)  + O(h^2)\, .
\end{eqnarray}
Using the above expression for the relevant Riemann components, we see that the interaction Hamiltonian in the co-moving frame is given by 
\begin{eqnarray}
H_{I} = \frac{m_e}{2}H_{pq}x^{p}x^{q} \, ,
\end{eqnarray}
with $H_{pq} = \Big(-\delta_{pq}\frac{\ddot{a}}{a} -\frac{\dot{a}}{a}\dot{h}_{pq} -\frac{1}{2} \ddot{h}_{pq}\Big).$ \\

Taking this perturbative Hamiltonian we consider the transition of the hydrogen from $\psi_{nlm}$ to $\psi_{n'l'm'}$ while the quantized gravitational perturbations makes a transition from an initial vacuum state $|0\rangle$ to any arbitrary final state. The transition probability for this case is given by \cite{dhanuka2022unruh}
\begin{eqnarray}\label{5}
P_{\psi_{nlm}\to \psi_{n'l'm'}} &=& \frac{m_e^2}{4}\bra {\psi_{n'l'm'}}\hat{x}^{i}\hat{x}^{j}\ket{\psi_{nlm}}^{*}\bra {\psi_{n'l'm'}}\hat{x}^{p}\hat{x}^{k}\ket{\psi_{nlm}}  \\  &\times & \int^{\eta_{f}}_{\eta_{i}}d\eta_{1} \int^{\eta_{f}}_{\eta_{i}}d\eta_{2}e^{-i\Omega(t({\eta_{1}})-t(\eta_{2}))}a(\eta_{1})a(\eta_{2})\bra{0} \hat{H}_{ij}(\vec{y},\eta_{1})\hat{H}_{pk}(\vec{y},\eta_{2})\ket{0} \, ,
\end{eqnarray}
where $\Omega = E_{n'l'}-E_{nl}$ is the energy difference between the two atomic states and $\vec{y}$ represents the fixed spatial coordinates for the co-moving trajectory the atom is moving along. In the absence of gravitational perturbations, $h_{lm}$, the interaction Hamiltonian reduces to  $H_{lm} = -\delta_{lm}\ddot{a}/a$, which forbids the transitions between eigenfunctions corresponding to different spherical harmonics, which is along expected lines due  to homogeneity and isotropy of the unperturbed background curvature. However, with gravitational waves turned on, we expect to have transitions with non-trivial change in spherical harmonics, which can be used as a distinct imprint of the quantum gravitons on the transitions of the hydrogen atom. We now show even the probability rate of such transitions is overwhelmingly large. Using the unperturbed flat spacetime hydrogen atom  energy eigenstates \cite{Griffiths1}, we can find the selection rules for the transitions of the form 
$<n',l',m'|x^{i}x^{p}|n,l,m>$, which are given in the table \ref{SelectionRules}. 
\begin{table}[h]
\centering
\begin{tabular}{|cc|cc|cc|}
\hline
\multicolumn{2}{|c|}{$x^2, y^2, xy, yx$}      & \multicolumn{2}{c|}{$xz, yz$}              & \multicolumn{2}{c|}{$z^2$}                   \\ \hline
\multicolumn{1}{|c|}{$\Delta l$} & $\Delta m$ & \multicolumn{1}{c|}{$\Delta l$} & $\Delta m$ & \multicolumn{1}{c|}{$\Delta l$} & $\Delta m$ \\ \hline
\multicolumn{1}{|c|}{-2,0,2}     & -2,0,2     & \multicolumn{1}{c|}{-2,0,2}     & -1,1       & \multicolumn{1}{c|}{-2,0,2}     & 0          \\ \hline
\end{tabular}
\caption{Selection rules for transitions of the type $<n',l',m'|x^{i}x^{p}|n,l,m>$.} \label{SelectionRules}
\end{table}
Further, using the equation of motion in Eq.(\ref{9}), we see that $H_{lm}$ can be written as 
\begin{equation}\label{1}
H_{pq} = -\frac{\delta_{pq}}{a^4}(aa'' - a'^2) + \frac{1}{a^4}\Big(\frac{aa'}{2}\frac{\partial}{\partial\eta} -\frac{a^2}{2} \nabla^{2}_{\vec{y}}\Big)h_{pq}(\eta,\vec{y}).
\end{equation}

\noindent Let us now consider a transition between different spherical harmonics. For these transitions, the unperturbed curvature induced term proportional to $\delta_{lm}$ in $H_{lm}$ drops out and we see that, using (\ref{1}), the vacuum expectation of the product of gravitational fields appearing in the formula for transition probability is given by
\begin{multline}\label{4}
\lim_{\vec{y}_1 \to \vec{y}_2}\bra{0} \hat{H}_{ij}(\vec{y}_{1},\eta_{1})\hat{H}_{pk}(\vec{y}_{2},\eta_{2})\ket{0} \\ = \lim_{\vec{y}_1 \to \vec{y}_2}\frac{1}{a_{1}^{4}}\Big(\frac{a_{1}a_{1}^{'}}{2}\frac{\partial}{\partial\eta_{1}} -\frac{a_{1}^{2}}{2} \nabla^{2}_{\vec{y}_{1}}\Big)\frac{1}{a_{2}^{4}}\Big(\frac{a_{2}a_{2}^{'}}{2}\frac{\partial}{\partial\eta_{2}} -\frac{a_{2}^2}{2} \nabla^{2}_{\vec{y}_{2}}\Big)\bra{0} \hat{h}_{ij}(\vec{y}_{1},\eta_1)\hat{h}_{pk}(\vec{y}_{2},\eta_2)\ket{0}.
\end{multline}
\noindent Using (\ref{8}), (\ref{5}) and (\ref{4}), we obtain the expression for transition probability between states with different spherical harmonics as
\begin{multline}\label{7}
P_{\psi_{nlm}\to \psi_{n'l'm'}} = \frac{m^2}{4}\bra {\psi_{n'l'm'}}\hat{x}^{i}\hat{x}^{j}\ket{\psi_{nlm}}^{*}\bra {\psi_{n'l'm'}}\hat{x}^{p}\hat{x}^{k}\ket{\psi_{nlm}}   \lim_{\vec{y}_1 \to \vec{y}_2}\int^{\eta_{f}}_{\eta_{i}}d\eta_{1}\int^{\eta_{f}}_{\eta_{i}}d\eta_{2}e^{-i\Omega(t(\eta_{1})-t(\eta_{2}))}\\ \frac{1}{a_{1}^{3}}\Big(\frac{a_{1}a_{1}^{'}}{2}\frac{\partial}{\partial\eta_{1}} -\frac{a_{1}^{2}}{2} \nabla^{2}_{\vec{y}_{1}}\Big)\frac{1}{a_{2}^{3}}\Big(\frac{a_{2}a_{2}^{'}}{2}\frac{\partial}{\partial\eta_{2}} -\frac{a_{2}^2}{2} \nabla^{2}_{\vec{y}_{2}}\Big)\Big(\delta_{ip}\delta_{jk} + \delta_{ik}\delta_{jp} - \delta_{ij}\delta_{pk} + \delta_{ij}\frac{\partial_{\vec{y}_{1p}}\partial_{\vec{y}_{1k}}}{\nabla_{\vec{y}_1}^{2}} + \delta_{pk}\frac{\partial_{\vec{y}_{1i}}\partial_{\vec{y}_{1j}}}{\nabla_{\vec{y}_1}^{2}}  -\delta_{ip}\frac{\partial_{\vec{y}_{1j}}\partial_{\vec{y}_{1k}}}{\nabla_{\vec{y}_1}^{2}} \\ - \delta_{ik}\frac{\partial_{\vec{y}_{1j}}\partial_{\vec{y}_{1p}}}{\nabla_{\vec{y}_1}^{2}}-\delta_
{jp}\frac{\partial_{\vec{y}_{1i}}\partial_{\vec{y}_{1k}}}{\nabla_{\vec{y}_1}^{2}}-\delta_{jk}\frac{\partial_{\vec{y}_{1i}}\partial_{\vec{y}_{1p}}}{\nabla_{\vec{y}_1}^{2}} + \frac{\partial_{\vec{y}_{1i}}\partial_{\vec{y}_{1j}}\partial_{\vec{y}_{1p}}\partial_{\vec{y}_{1k}}}{\nabla_{\vec{y}_1}^{2}\nabla_{\vec{y}_1}^{2}}\Big)\underbrace{\int d^3\vec{q} e^{i\vec{q}.(\vec{y}_{1}-\vec{y}_{2})}h_{q}(\eta_{1})h^{*}_{q}(\eta_{2})}_{G_{\text{Matter}}(x(\eta_{1}),x(\eta_{2}))} \, .
\end{multline}
We can now go to a new set of variables $\tilde{\eta} = \frac{\eta_{1} + \eta_{2}}{2}$ and $\Delta \eta = \eta_{1}-\eta_{2}$ and define the rate of transitions with respect to the variable $\tilde{\eta}$. In fact, one obtains the following expression for the rate with respect to $\tilde{\eta}$
\begin{multline}\label{10}
\frac{dP_{\psi_{nlm}\to \psi_{n'l'm'}}}{d\tilde{\eta}} = \frac{m^2}{4}\bra {\psi_{n'l'm'}}\hat{x}^{i}\hat{x}^{j}\ket{\psi_{nlm}}^{*}\bra {\psi_{n'l'm'}}\hat{x}^{p}\hat{x}^{k}\ket{\psi_{nlm}}   \lim_{\vec{y}_1 \to \vec{y}_2}\int^{2(\tilde{\eta}-\eta_{i})}_{-2(\tilde{\eta}-\eta_{i})}d(\Delta\eta)\Bigg[e^{-i\Omega(t(\eta_{1})-t(\eta_{2}))}\\ \frac{1}{a_{1}^{3}}\Big(\frac{a_{1}a_{1}^{'}}{2}\frac{\partial}{\partial\eta_{1}} -\frac{a_{1}^{2}}{2} \nabla^{2}_{\vec{y}_{1}}\Big)\frac{1}{a_{2}^{3}}\Big(\frac{a_{2}a_{2}^{'}}{2}\frac{\partial}{\partial\eta_{2}} -\frac{a_{2}^2}{2} \nabla^{2}_{\vec{y}_{2}}\Big)\Big(\delta_{ip}\delta_{jk} + \delta_{ik}\delta_{jp} - \delta_{ij}\delta_{pk} + \delta_{ij}\frac{\partial_{\vec{y}_{1p}}\partial_{\vec{y}_{1k}}}{\nabla_{\vec{y}_1}^{2}} + \delta_{pk}\frac{\partial_{\vec{y}_{1i}}\partial_{\vec{y}_{1j}}}{\nabla_{\vec{y}_1}^{2}}  -\delta_{ip}\frac{\partial_{\vec{y}_{1j}}\partial_{\vec{y}_{1k}}}{\nabla_{\vec{y}_1}^{2}} \\ - \delta_{ik}\frac{\partial_{\vec{y}_{1j}}\partial_{\vec{y}_{1p}}}{
\nabla_{\vec{y}_1}^{2}}-\delta_{jp}\frac{\partial_{\vec{y}_{1i}}\partial_{\vec{y}_{1k}}}{\nabla_{\vec{y}_1}^{2}}-\delta_{jk}\frac{\partial_{\vec{y}_{1i}}\partial_{\vec{y}_{1p}}}{\nabla_{\vec{y}_1}^{2}} + \frac{\partial_{\vec{y}_{1i}}\partial_{\vec{y}_{1j}}\partial_{\vec{y}_{1p}}\partial_{\vec{y}_{1k}}}{\nabla_{\vec{y}_1}^{2}\nabla_{\vec{y}_1}^{2}}\Big)\underbrace{\int d^3\vec{q} e^{i\vec{q}.(\vec{y}_{1}-\vec{y}_{2})}h_{q}(\eta_{1})h^{*}_{q}(\eta_{2})}_{G_{\text{Matter}}(x(\eta_{1}),x(\eta_{2}))}\Bigg] \, ,
\end{multline}
where the quantity in the big square brackets is first evaluated as given and then every factor of $\eta_{1}$ and $\eta_{2}$ in it is to be expressed in terms of $\tilde{\eta}$ and $\Delta \eta$ before performing the $\Delta \eta$ integral.
\section{Comparison with Electromagnetic transition}

\noindent Let us consider the action of Maxwell theory in some spacetime with metric $g_{\mu\nu}$ i.e., 
\begin{equation}
S[A_{\alpha}, g_{\mu\nu}] = -\frac{1}{4} \int d^{4} x \sqrt{-g} F_{\mu\nu} F_{\rho \sigma} g^{\mu \rho}g^{\nu \sigma} \, .
\end{equation}
If we consider another system where we have 
\begin{eqnarray}
\tilde{g}_{\mu\nu} &=& a^{2}g_{\mu \nu} \, ,\nonumber \\
\tilde{A}_{\rho} &=& A_{\rho} \, ,\nonumber \\
\tilde{g}^{\mu\nu} &=& a^{-2}g^{\mu \nu}\, , \nonumber \\
\sqrt{-\tilde{g}} &=& a^{4}\sqrt{-g} \, ,
\end{eqnarray}
we find that 
\begin{eqnarray}
\tilde{S}[\tilde{A}_{\alpha}, \tilde{g}_{\mu\nu}] &=& -\frac{1}{4} \int d^{4} x \sqrt{-\tilde{g}} \tilde{F}_{\mu\nu} \tilde{F}_{\rho \sigma} \tilde{g}^{\mu \rho}\tilde{g}^{\nu \sigma} \, , \nonumber \\
&=& -\frac{1}{4} \int d^{4} x \sqrt{-g}a^{4}F_{\mu\nu} F_{\rho \sigma} a^{-2} g^{\mu \rho}a^{-2} g^{\nu \sigma}\, , \nonumber \\
&=& S[A_{\alpha}, g_{\mu\nu}] \, .
\end{eqnarray}

Some important relations are 
\begin{eqnarray}
\tilde{A}^{\mu} &=& \tilde{g}^{\mu\nu}\tilde{A}_{\nu} = a^{-2}g^{\mu\nu}A_{\nu} = a^{-2}A^{\mu} \, ,\nonumber \\
\tilde{F}^{\mu \nu} &=& a^{-4}F^{\mu\nu} \, .
\end{eqnarray}

Particularly, we have 
\begin{equation}
\tilde{E}^{i} = \tilde{F}^{0i} = a^{-4}F^{0i} = a^{-4}E^{i} \, .
\end{equation}

Specializing to the case where we have $g_{\mu\nu} = \eta_{\mu\nu}$ i.e., $ds^2= -(d\eta)^2 + (d\vec{x})^2$, and $\tilde{g}_{\mu\nu} = a^2 \eta_{\mu\nu}$, we have  
\begin{equation}
\tilde{E}^{i}(x) = a^{-4}(\eta)E^{i}(x) = -a^{-4}(\eta)\partial_{\eta}A^{i}(x) \, ,
\end{equation}
in the Coulomb gauge in which $A^{0}(x) = 0$ and $\vec{\nabla}.\vec{A} = 0$. \\ 
This implies that 
\begin{eqnarray}
\langle \tilde{E}^{i}(x_1)\tilde{E}^{j}(x_2)\rangle &=& a^{-4}_{1}a^{-4}_{2} \partial_{\eta_{1}}\partial_{\eta_{2}} \langle A^{i}(x_1)A^{j}(x_2)\rangle \, , \nonumber \\
&=&  a^{-4}_{1}a^{-4}_{2} (\partial_{\eta_{1}}\partial_{\eta_{2}}\delta^{ij} - \partial^{i}_{1}\partial^{j}_{2})  \frac{1}{4\pi^2} \Big(\frac{1}{-(\Delta \eta - i\epsilon)^2 + (\Delta \vec{x})^2}\Big)\, , \nonumber \\
& = & \frac{a^{-4}_{1}a^{-4}_{2}\Big((-(\Delta \eta - i\epsilon)^2 - (\Delta \vec{x})^2)\delta_{ij} + 2(x_{1}-x_{2})_{i}(x_{1}-x_{2})_{j} \Big)}{\pi^2(-(\Delta \eta - i\epsilon)^2 + (\Delta \vec{x})^2)^3} \, .
\end{eqnarray}
\noindent For comoving observers, the above expression reduces to 
\begin{equation}
\langle \tilde{E}^{i}(x_1)\tilde{E}^{j}(x_2)\rangle=\frac{a^{-4}_{1}a^{-4}_{2} \delta_{ij} }{\pi^2(\Delta \eta - i\epsilon)^4 } \, .
\end{equation}
\noindent The dipole interaction is given by 
\begin{equation}
H_{I} = - d_{\mu}E^{\mu} = - d_{\mu}F^{\mu\nu}u_{\nu} \, ,
\end{equation}
where $u_{\nu} = (-1,0,0,0)$ in the comoving coordinates. Thus, we obtain that 
\begin{equation}
H_{I} = d_{\mu}F^{\mu 0} = -d_{i}F^{0i} = -d_{i}E^{i}_{comoving} \, .
\end{equation}
But $E^{i}_{comoving} = F^{0i}_{comoving} = (\partial t)/(\partial \eta) F^{0i}_{conformal} = a(\eta) E^{i}_{conformal}.$ We also have $E^{i}_{conformal} = a^{-4}E^{i}_{flat}.$ This implies that 
\begin{equation}
H_{I} =  -d_{i}a(\eta)E^{i}_{conformal} = - a^{-3}d_{i}E^{i}_{flat} = - a^{-1}d^{i}E^{i}_{flat} \, .
\end{equation}
The probability for a UdW detector to make a transition from ground state to excited state , within first order perturbation theory, is given by 
\begin{eqnarray}
\frac{P_{0 \to \Omega}}{|{}_{D}\bra{\Omega}\hat{d}^{z}(0)\ket{0}_{D}|^2} &=& \int_{t_{i}}^{t_{f}} \int_{t_{i}}^{t_{f}} dt_{1} dt_{2}e^{-i\Omega(t_{1}-t_{2})}a^{-1}(t_{1})a^{-1}(t_{2}) \langle \tilde{E}^{z}_{flat}(x_1)\tilde{E}^{z}_{flat}(x_2)\rangle\, ,\nonumber \\ &= & \int_{\eta_{i}}^{\eta_{f}} \int_{\eta_{i}}^{\eta_{f}} d\eta_{1} d\eta_{2}e^{-i\Omega(t(\eta_{1})-t(\eta_{2}))} \langle \tilde{E}^{z}_{flat}(x_1)\tilde{E}^{z}_{flat}(x_2)\rangle\, , \nonumber \\ &= & \int_{\eta_{i}}^{\eta_{f}} \int_{\eta_{i}}^{\eta_{f}} d\eta_{1} d\eta_{2}e^{-i\Omega(t(\eta_{1})-t(\eta_{2}))} \frac{1}{\pi^2 (\Delta \eta - i\epsilon)^4}\, .
\end{eqnarray}
Let us consider the spacetimes with $q \in (-2,0)$ for which the cosmic time is related to the conformal time by the relation $t = \frac{H^{-q}\eta^{1-q}}{1-q}$ and $t \in (0, \infty)$ for $\eta \in (0, \infty)$. Defining a new variable $z = \Omega H^{-q}\eta^{1-q}$, we can pull out all the $\Omega$ and $H$ dependence out of the above integral  as
\begin{eqnarray}
\frac{P_{0 \to \Omega}}{|{}_{D}\bra{\Omega}\hat{d}^{z}(0)\ket{0}_{D}|^2}=  \int_{z_{i}}^{z_{f}} \int_{z_{i}}^{z_{f}}\underbrace{(\Omega H^{-q})^{\frac{-1}{1-q}} \frac{ dz_{1}}{\left[(1-q)z_1\right]^{\frac{-q}{1-q}}}}_{d\eta_1}\underbrace{ (\Omega H^{-q})^{\frac{-1}{1-q}}\frac{ dz_{2}}{\left[(1-q)z_2\right]^{\frac{-q}{1-q}}}}_{d\eta_2} \left[\frac{\Omega H^{-q}}{1-q}\right]^{\frac{4}{1-q}}   \frac{ e^{-i(z_1-z_2)}}{\pi^2 \left(z_1^{\frac{1}{1-q}} -z_2^{\frac{1}{1-q}}- i\epsilon\right)^4}\nonumber. 
\end{eqnarray}
The characterstic rate of transition probability, $\Gamma_{e}$, through the duilution of one $\eta$ integral  is obtained as
\begin{equation}\label{RRE}
\Gamma_{e} \propto d^2 (\Omega H^{-q})^{\frac{3}{1-q}} \, ,
\end{equation}
where $d^2 = |{}_{D}\bra{\Omega}\hat{d}^{z}(0)\ket{0}_{D}|^2$
is the dipole moment square. The above expression for rate considers the case in which the limits for $\eta_{1}$ and $\eta_{2}$ are taken to be $(0,\infty)$.   
\section{Response rate for gravitational waves in curved spacetime}
\noindent The transition probability between states with different spherical harmonics of a hydrogen atom caused by gravitational waves is given by eq.~\ref{7} i.e., 
\begin{multline}
P_{\psi_{nlm}\to \psi_{n'l'm'}} = \frac{m_e^2}{4}\bra {\psi_{n'l'm'}}\hat{x}^{i}\hat{x}^{j}\ket{\psi_{nlm}}^{*}\bra {\psi_{n'l'm'}}\hat{x}^{p}\hat{x}^{k}\ket{\psi_{nlm}}   \lim_{\vec{c}_1 \to \vec{c}_2}\int^{\eta_{f}}_{\eta_{i}}d\eta_{1}\int^{\eta_{f}}_{\eta_{i}}d\eta_{2}e^{-i\Omega(t(\eta_{1})-t(\eta_{2}))}\\ \frac{1}{a_{1}^{3}}\Big(\frac{a_{1}a_{1}^{'}}{2}\frac{\partial}{\partial\eta_{1}} -\frac{a_{1}^{2}}{2} \nabla^{2}_{\vec{c}_{1}}\Big)\frac{1}{a_{2}^{3}}\Big(\frac{a_{2}a_{2}^{'}}{2}\frac{\partial}{\partial\eta_{2}} -\frac{a_{2}^2}{2} \nabla^{2}_{\vec{c}_{2}}\Big)\Big(\delta_{ip}\delta_{jk} + \delta_{ik}\delta_{jp} - \delta_{ij}\delta_{pk} + \delta_{ij}\frac{\partial_{\vec{c}_{1p}}\partial_{\vec{c}_{1k}}}{\nabla_{\vec{c}_1}^{2}} + \delta_{pk}\frac{\partial_{\vec{c}_{1i}}\partial_{\vec{c}_{1j}}}{\nabla_{\vec{c}_1}^{2}}  -\delta_{ip}\frac{\partial_{\vec{c}_{1j}}\partial_{\vec{c}_{1k}}}{\nabla_{\vec{c}_1}^{2}} \\ - \delta_{ik}\frac{\partial_{\vec{c}_{1j}}\partial_{\vec{c}_{1p}}}{\nabla_{\vec{c}_1}^{2}}-\delta_
{jp}\frac{\partial_{\vec{c}_{1i}}\partial_{\vec{c}_{1k}}}{\nabla_{\vec{c}_1}^{2}}-\delta_{jk}\frac{\partial_{\vec{c}_{1i}}\partial_{\vec{c}_{1p}}}{\nabla_{\vec{c}_1}^{2}} + \frac{\partial_{\vec{c}_{1i}}\partial_{\vec{c}_{1j}}\partial_{\vec{c}_{1p}}\partial_{\vec{c}_{1k}}}{\nabla_{\vec{c}_1}^{2}\nabla_{\vec{c}_1}^{2}}\Big)\int d^3\vec{q} e^{i\vec{q}.(\vec{c}_{1}-\vec{c}_{2})}h_{q}(\eta_{1})h^{*}_{q}(\eta_{2}) \, .
\end{multline}
Now considering transition between those states for which the matrix elements involved in the above expression are non-zero and pulling out all the $\Omega$ and $H$ dependence out of the above integral, just as done for the electromagnetic case above, we find that the rate of transition, $\Gamma_{g}$, caused by gravitational waves is
\begin{equation}\label{RRG}
\Gamma_{g} \propto m_{e} ^2 (Q)^2 (\Omega)^{\frac{5-4q}{1-q}} (H)^{\frac{-q}{1-q}} \, ,
\end{equation}
where $Q = \bra {\psi_{n'l'm'}}\hat{x}^{z}\hat{x}^{z}\ket{\psi_{nlm}} $ or any other such matrix elements. In this case also, the limits for $\eta_{1}$ and $\eta_{2}$ are taken to be $(0,\infty)$. 
\section{Flat spacetime case}
\noindent Using eqs.~\ref{RRE} and \ref{RRG} and taking $q=0$, we find that the ratio of response rate caused by electromagnetic and gravitational fields in flat spacetime i.e., for $ q = 0$, is given by 
\begin{equation}
\Big(\frac{\Gamma_{e}}{\Gamma_{g}}\Big)_{flat} \propto \frac{d^2}{m_{e} ^2 (Q)^2 \Omega ^2} \, .
\end{equation}
\section{Curved spacetime case}
\noindent In curved spacetime, the ratio of rates has the following form 
\begin{equation}
\Big(\frac{\Gamma_{e}}{\Gamma_{g}}\Big)_{curved} \propto \frac{d^2}{m_{e} ^2 (Q)^2 \Omega ^2}  \Big(\frac{H}{\Omega}\Big)^{-\frac{2q}{1-q}}  \, .
\end{equation}

Furthermore, in a matter driven universe, the graviton correlator function has a divergent piece as suggested, thus the ratio further swings in the favor of gravitational transitions through the parameter $\delta =2+q \rightarrow 0$
\begin{equation}
\Big(\frac{\Gamma_{e}}{\Gamma_{g}}\Big)_{curved} \propto \delta \frac{d^2}{m_{e} ^2 (Q)^2 \Omega ^2}  \Big(\frac{H}{\Omega}\Big)^{-\frac{2q}{1-q}}  \, .
\end{equation}


\begin{thebibliography}{10}
\bibitem{Kiefer:2005uk}
Claus Kiefer.
\newblock {Quantum gravity: General introduction and recent developments}.
\newblock {\em Annalen Phys.}, 15:129--148, 2005.

\bibitem{Boughn:2006st}
Stephen Boughn and Tony Rothman.
\newblock {Aspects of graviton detection: Graviton emission and absorption by
  atomic hydrogen}.
\newblock {\em Class. Quant. Grav.}, 23:5839--5852, 2006.

\bibitem{scully_zubairy_1997}
Marlan~O. Scully and M.~Suhail Zubairy.
\newblock {\em Quantum Optics}.
\newblock Cambridge University Press, 1997.

\bibitem{milonni1994quantum}
P.W. Milonni.
\newblock {\em The Quantum Vacuum: An Introduction to Quantum Electrodynamics}.
\newblock Elsevier Science, 1994.

\bibitem{loudon2000quantum}
R.~Loudon.
\newblock {\em The Quantum Theory of Light}.
\newblock OUP Oxford, 2000.

\bibitem{Martin-Martinez:2012ysv}
Eduardo Martin-Martinez, Miguel Montero, and Marco del Rey.
\newblock {Wavepacket detection with the Unruh-DeWitt model}.
\newblock {\em Phys. Rev. D}, 87(6):064038, 2013.

\bibitem{Guerreiro:2019vbq}
Thiago Guerreiro.
\newblock {Quantum Effects in Gravity Waves}.
\newblock {\em Class. Quant. Grav.}, 37(15):155001, 2020.

\bibitem{Parikh:2020nrd}
Maulik Parikh, Frank Wilczek, and George Zahariade.
\newblock {The Noise of Gravitons}.
\newblock {\em Int. J. Mod. Phys. D}, 29(14):2042001, 2020.

\bibitem{Anastopoulos:2020cdp}
Charis Anastopoulos and Bei-Lok Hu.
\newblock {Quantum Superposition of Two Gravitational Cat States}.
\newblock {\em Class. Quant. Grav.}, 37(23):235012, 2020.

\bibitem{Parikh:2020kfh}
Maulik Parikh, Frank Wilczek, and George Zahariade.
\newblock {Quantum Mechanics of Gravitational Waves}.
\newblock {\em Phys. Rev. Lett.}, 127(8):081602, 2021.

\bibitem{Parikh:2020fhy}
Maulik Parikh, Frank Wilczek, and George Zahariade.
\newblock {Signatures of the quantization of gravity at gravitational wave
  detectors}.
\newblock {\em Phys. Rev. D}, 104(4):046021, 2021.

\bibitem{Marletto:2017kzi}
Chiara Marletto and Vlatko Vedral.
\newblock {Gravitationally-induced entanglement between two massive particles
  is sufficient evidence of quantum effects in gravity}.
\newblock {\em Phys. Rev. Lett.}, 119(24):240402, 2017.

\bibitem{Bose:2017nin}
Sougato Bose, Anupam Mazumdar, Gavin~W. Morley, Hendrik Ulbricht, Marko
  Toro\v{s}, Mauro Paternostro, Andrew Geraci, Peter Barker, M.~S. Kim, and
  Gerard Milburn.
\newblock {Spin Entanglement Witness for Quantum Gravity}.
\newblock {\em Phys. Rev. Lett.}, 119(24):240401, 2017.

\bibitem{krisnanda2020observable}
Tanjung Krisnanda, Guo~Yao Tham, Mauro Paternostro, and Tomasz Paterek.
\newblock Observable quantum entanglement due to gravity.
\newblock {\em npj Quantum Information}, 6(1):1--6, 2020.

\bibitem{Belenchia:2018szb}
Alessio Belenchia, Robert~M. Wald, Flaminia Giacomini, Esteban Castro-Ruiz,
  \v{C}aslav Brukner, and Markus Aspelmeyer.
\newblock {Quantum Superposition of Massive Objects and the Quantization of
  Gravity}.
\newblock {\em Phys. Rev. D}, 98(12):126009, 2018.

\bibitem{Carney:2018ofe}
Daniel Carney, Philip C.~E. Stamp, and Jacob~M. Taylor.
\newblock {Tabletop experiments for quantum gravity: a user\textquoteright{}s
  manual}.
\newblock {\em Class. Quant. Grav.}, 36(3):034001, 2019.

\bibitem{Kanno:2021gpt}
Sugumi Kanno, Jiro Soda, and Junsei Tokuda.
\newblock {Indirect detection of gravitons through quantum entanglement}.
\newblock {\em Phys. Rev. D}, 104(8):083516, 2021.

\bibitem{Rothman:2006fp}
Tony Rothman and Stephen Boughn.
\newblock {Can gravitons be detected?}
\newblock {\em Found. Phys.}, 36:1801--1825, 2006.

\bibitem{Jones:2015uda}
Preston Jones and Douglas Singleton.
\newblock {Gravitons to photons \textemdash{} Attenuation of gravitational
  waves}.
\newblock {\em Int. J. Mod. Phys. D}, 24(12):1544017, 2015.

\bibitem{Martin:2000xs}
Jerome Martin and Robert~H. Brandenberger.
\newblock {The TransPlanckian problem of inflationary cosmology}.
\newblock {\em Phys. Rev. D}, 63:123501, 2001.

\bibitem{Damour:1995pd}
Thibault Damour and Alexander Vilenkin.
\newblock {String theory and inflation}.
\newblock {\em Phys. Rev. D}, 53:2981--2989, 1996.

\bibitem{Tsujikawa:2003vr}
Shinji Tsujikawa, Parampreet Singh, and Roy Maartens.
\newblock {Loop quantum gravity effects on inflation and the CMB}.
\newblock {\em Class. Quant. Grav.}, 21:5767--5775, 2004.

\bibitem{Bojowald:2002nz}
Martin Bojowald.
\newblock {Inflation from quantum geometry}.
\newblock {\em Phys. Rev. Lett.}, 89:261301, 2002.

\bibitem{Kiefer:2008ku}
Claus Kiefer and David Polarski.
\newblock {Why do cosmological perturbations look classical to us?}
\newblock {\em Adv. Sci. Lett.}, 2:164--173, 2009.

\bibitem{Brandenberger:1990bx}
Robert~H. Brandenberger, Raymond Laflamme, and Milan Mijic.
\newblock {Classical Perturbations From Decoherence of Quantum Fluctuations in
  the Inflationary Universe}.
\newblock {\em Mod. Phys. Lett. A}, 5:2311--2318, 1990.

\bibitem{CMBPolStudyTeam:2008rgp}
Daniel Baumann et~al.
\newblock {CMBPol Mission Concept Study: Probing Inflation with CMB
  Polarization}.
\newblock {\em AIP Conf. Proc.}, 1141(1):10--120, 2009.

\bibitem{Parker:2009uva}
Leonard~E. Parker and D.~Toms.
\newblock {\em {Quantum Field Theory in Curved Spacetime}: {Quantized Field and
  Gravity}}.
\newblock Cambridge Monographs on Mathematical Physics. Cambridge University
  Press, 8 2009.

\bibitem{Weinberg:2008zzc}
S.~Weinberg.
\newblock {\em Cosmology}.
\newblock Cosmology. Oxford University Press, Oxford, 2008.

\bibitem{Caprini:2018mtu}
Chiara Caprini and Daniel~G. Figueroa.
\newblock {Cosmological Backgrounds of Gravitational Waves}.
\newblock {\em Class. Quant. Grav.}, 35(16):163001, 2018.

\bibitem{Parker:1968mv}
L.~Parker.
\newblock {Particle creation in expanding universes}.
\newblock {\em Phys. Rev. Lett.}, 21:562--564, 1968.

\bibitem{Parker:1969au}
Leonard Parker.
\newblock {Quantized fields and particle creation in expanding universes. 1.}
\newblock {\em Phys. Rev.}, 183:1057--1068, 1969.

\bibitem{Birrell:1982ix}
N.~D. Birrell and P.~C.~W. Davies.
\newblock {\em {Quantum Fields in Curved Space}}.
\newblock Cambridge Monographs on Mathematical Physics. Cambridge Univ. Press,
  Cambridge, UK, 2 1984.

\bibitem{Mottola:1984ar}
E.~Mottola.
\newblock {Particle Creation in de Sitter Space}.
\newblock {\em Phys. Rev. D}, 31:754, 1985.

\bibitem{Lochan:2018pzs}
Kinjalk Lochan, Karthik Rajeev, Amit Vikram, and T.~Padmanabhan.
\newblock {Quantum correlators in Friedmann spacetimes: The omnipresent de
  Sitter spacetime and the invariant vacuum noise}.
\newblock {\em Phys. Rev. D}, 98(10):105015, 2018.

\bibitem{Dhanuka:2020yxp}
Ankit Dhanuka and Kinjalk Lochan.
\newblock {Stress energy correlator in de Sitter spacetime: Its conformal
  masking or growth in connected Friedmann universes}.
\newblock {\em Phys. Rev. D}, 102(8):085009, 2020.

\bibitem{Lochan:2022dht}
Kinjalk Lochan.
\newblock {Unequal time commutators in Friedmann universes: deterministic
  evolution of massless fields}.
\newblock {\em Gen. Rel. Grav.}, 54(9):100, 2022.

\bibitem{White:1994bn}
Simon D.~M. White.
\newblock {Formation and evolution of galaxies: Lectures given at Les Houches,
  August 1993}.
\newblock In {\em {Les Houches Summer School on Cosmology and Large Scale
  Structure (Session 60)}}, pages 349--430, 8 1994.
  
\bibitem{Lochan:2014xja}
Kinjalk Lochan and T.~Padmanabhan.
\newblock {Inertial nonvacuum states viewed from the Rindler frame}.
\newblock {\em Phys. Rev. D}, 91(4):044002, 2015.
  

\bibitem{Allen:1987tz}
Bruce Allen and Antoine Folacci.
\newblock {The Massless Minimally Coupled Scalar Field in De Sitter Space}.
\newblock {\em Phys. Rev. D}, 35:3771, 1987.

\bibitem{dhanuka2022unruh}
Ankit Dhanuka and Kinjalk Lochan.
\newblock {Unruh DeWitt probe of late time revival of quantum correlations in
  Friedmann spacetimes}.
\newblock {\em Phys. Rev. D}, 106:125006, 2022.

\bibitem{Stargen:2021vtg}
D.~Jaffino Stargen and Kinjalk Lochan.
\newblock {Cavity Optimization for Unruh Effect at Small Accelerations}.
\newblock {\em Phys. Rev. Lett.}, 129(11):111303, 2022.

\bibitem{Kolb:1990vq}
Edward~W. Kolb and Michael~S. Turner.
\newblock {\em {The Early Universe}}, volume~69.
\newblock 1990.

\bibitem{Parker:1980kw}
L.~Parker.
\newblock {ONE ELECTRON ATOM AS A PROBE OF SPACE-TIME CURVATURE}.
\newblock {\em Phys. Rev. D}, 22:1922--1934, 1980.

\bibitem{Parker:1980hlc}
Leonard Parker.
\newblock {One-Electron Atom in Curved Space-Time}.
\newblock {\em Phys. Rev. Lett.}, 44(23):1559, 1980.

\bibitem{Parker:1982nk}
L.~Parker and L.~O. Pimentel.
\newblock {GRAVITATIONAL PERTURBATION OF THE HYDROGEN SPECTRUM}.
\newblock {\em Phys. Rev. D}, 25:3180--3190, 1982.

\bibitem{bessis1984}
N.~Bessis, G.~Bessis, and R.~Shamseddine.
\newblock Space-curvature effects in atomic fine- and hyperfine-structure
  calculations.
\newblock {\em Phys. Rev. A}, 29:2375--2388, May 1984.

\bibitem{Poisson:2009pwt}
Eric Poisson.
\newblock {\em {A Relativist's Toolkit: The Mathematics of Black-Hole
  Mechanics}}.
\newblock Cambridge University Press, 12 2009.

\bibitem{Juarez-Aubry:2014jba}
Benito~A. Ju\'arez-Aubry and Jorma Louko.
\newblock {Onset and decay of the 1 + 1 Hawking-Unruh effect: what the
  derivative-coupling detector saw}.
\newblock {\em Class. Quant. Grav.}, 31(24):245007, 2014.

\bibitem{Juarez-Aubry:2018ofz}
Benito~A. Ju\'arez-Aubry and Jorma Louko.
\newblock {Quantum fields during black hole formation: How good an
  approximation is the Unruh state?}
\newblock {\em JHEP}, 05:140, 2018.


\bibitem{DiValentino:2021izs}
E.~Di Valentino, O.~Mena, S.~Pan, L.~Visinelli, W.~Yang, A.~Melchiorri, D.~F.~Mota, A.~G.~Riess and J.~Silk.
\newblock {\em In the realm of the Hubble tension\textemdash{}a review of solutions}.
\newblock {Class. Quant. Grav.} \textbf{38}, no.15, 153001, 2021.


\bibitem{Bowman:2018yin}
J.~D.~Bowman, A.~E.~E.~Rogers, R.~A.~Monsalve, T.~J.~Mozdzen and N.~Mahesh.
\newblock {\em An absorption profile centred at 78 megahertz in the sky-averaged spectrum}.
\newblock {Nature} \textbf{555}  no.7694, 67-70, 2018.

\bibitem{Brandenberger:1990bx}
Robert~H. Brandenberger, Raymond Laflamme, and Milan Mijic.
\newblock {Classical Perturbations From Decoherence of Quantum Fluctuations in
  the Inflationary Universe}.
\newblock {\em Mod. Phys. Lett. A}, 5:2311--2318, 1990.


\bibitem{Weinberg:2008zzc}
S.~Weinberg.
\newblock {\em Cosmology}.
\newblock Cosmology. Oxford University Press, Oxford, 2008.

\bibitem{Dai:2015rda}
Liang Dai, Enrico Pajer, and Fabian Schmidt.
\newblock {Conformal Fermi Coordinates}.
\newblock {\em JCAP}, 11:043, 2015.

\bibitem{dhanuka2022unruh}
Ankit Dhanuka and Kinjalk Lochan.
\newblock {Unruh DeWitt probe of late time revival of quantum correlations in
  Friedmann spacetimes}.
\newblock {\em Phys. Rev. D}, 106:125006, 2022.

\bibitem{Griffiths1}
David~J. Griffiths and Darrell~F. Schroeter.
\newblock {\em Introduction to Quantum Mechanics}.
\newblock Cambridge University Press, 3 edition, 2018.

\end{thebibliography}


\end{document}